\pdfoutput=1
\documentclass{article}
\usepackage{spconf,amsmath,graphicx,xcolor,hyperref}

\usepackage{amssymb}
\usepackage{booktabs}
\usepackage{multirow}

\newcommand{\coll}[1]{\textcolor{gray}{#1}}

\title{Collapse, Not Invariance: Diagnosing Auxiliary Objectives
       in Speech Anti-Spoofing}

\name{Ksenia Lysikova$^{1}$, Kirill Borodin$^{1,2}$, Maxim Maslov$^{1}$, Grach Mkrtchian$^{2,3}$}
\address{$^{1}$Lab260, Yerevan, Armenia \quad
         $^{2}$BitmanagerAI, Dubai, UAE \quad
         $^{3}$MTUCI, Moscow, Russia\\
         kborodin.research@gmail.com}

\makeatletter
\def\ps@ieeenotice{\let\@mkboth\@gobbletwo
  \def\@oddhead{}\def\@evenhead{}%
  \def\@oddfoot{\parbox[t]{\textwidth}{\centering\footnotesize
    This work has been submitted to the IEEE ICASSP for possible publication.
    Copyright may be transferred without notice, after which this version
    may no longer be accessible.}}%
  \let\@evenfoot\@oddfoot}
\makeatother

\begin{document}
\ninept
\maketitle
\thispagestyle{ieeenotice}

\begin{abstract}
Speech anti-spoofing countermeasures degrade when the generator, codec or
channel changes, and a common remedy is an auxiliary objective that shapes the
embedding space; whether it does is invisible to EER, a pure ranking metric.
We compare seven such objectives with cross-entropy over $113$ runs on five
corpora, AASIST3 at three seeds plus four pre-trained detectors, and measure
the embedding space of the $24$ AASIST3 runs directly. Raw augmentation
displacement makes cosine consistency look effective, but the gain is a
smaller space, not a more stable one: normalised by the spread, no
configuration consistently improves on cross-entropy. Every trained space is
dominated by the single decision axis expected for two classes, whose
training-set structure does not transfer, and four runs collapse to a
near-constant output that displacement rewards and EER reports as poor
accuracy. No auxiliary objective keeps an advantage over cross-entropy across
architectures, corpora and seeds.

\end{abstract}


\section{Introduction}
\label{sec:introduction}

Neural speech synthesis now produces speech hard to tell from genuine
recordings, and countermeasures near-perfect
in domain lose most of their accuracy once the generator, codec or channel
changes~\cite{asvspoof2021_yamagishi,muller22_generalize,muller24_harder}.

A common response, on top of a self-supervised
front-end~\cite{tak2022,viakhirev2025scalable}, is an auxiliary term on the
\emph{embedding space}, on the assumption that a better-structured
representation transfers more
readily~\cite{clad,attn_contrastive,grl_spoof}. These terms fall into four
families: \emph{consistency regularisation} pulls augmented views of an
utterance together~\cite{simsiam}, \emph{contrastive} learning also pushes
other utterances apart~\cite{simclr}, \emph{adversarial unlearning} strips
nuisance information~\cite{dann}, and \emph{label smoothing}, which adds no
term on the embedding but is known to compress the penultimate
geometry~\cite{mueller2019}, serves as a classifier-side control. Margin and
one-class objectives, OC-Softmax~\cite{zhang2021ocsoftmax} and
SAMO~\cite{ding2023samo}, and augmentation alone~\cite{rawboost} are further
routes to which the diagnostics below apply unchanged.

All rest on the premise that such a term shapes the space as intended and that
this structure is what transfers. EER cannot establish this: it depends only
on the score ranking, so it cannot tell a space shaped as intended from one
that was not, nor a collapsed encoder from a trained one that merely performs
poorly. We read the detector's internals instead, as~\cite{viakhirev2026interpret}
does for the branches of AASIST3. Each objective
is also typically reported on its own architecture and corpus, often from a
single run, a practice known to overstate differences between
systems~\cite{bouthillier2021variance}.

We compare seven auxiliary configurations with cross-entropy under one
protocol: AASIST3 at three seeds and four pre-trained detectors of different
design in three adaptation modes, $113$ runs evaluated on five corpora, with
the embedding space of the $24$ AASIST3 runs measured directly. Our
contributions are:

\begin{itemize}
  \item \textbf{A scale-aware protocol for embedding objectives.} Raw
  displacement $A$ changes under a rescaling that leaves the scores fixed, so
  we read it against the spread $S$ between utterances, confirm a vanishing
  $S$ with scale-free class separation, plot at a shared scale, and report
  seed-paired differences from cross-entropy.
  \item \textbf{Under it, apparent invariance is shrinkage.} The $39\%$ lower
  displacement of cosine consistency is a $42\%$ smaller space; normalised,
  no objective is more invariant than cross-entropy on every seed, and every
  trained run keeps the single decision axis cross-entropy produces for two
  classes~\cite{papyan2020nc}, whose training-set structure does not carry
  to evaluation data.
  \item \textbf{EER hides two failures the protocol exposes.} Four of $24$
  runs, from three objectives including one with no term on the embedding,
  collapse to a near-constant encoder that EER reports as poor accuracy and
  t-SNE draws as a healthy cloud; and single-seed rankings reverse:
  \texttt{MSE} beats cross-entropy by $12\%$ at one seed and trails every run
  at another.
\end{itemize}

\section{Objectives and Experimental Setup}
\label{sec:setup}

\subsection{Auxiliary objectives}
\label{sec:method_obj}

The primary backbone is the AASIST3 architecture~\cite{aasist3}, trained on
ASVspoof~5 from a pre-trained wav2vec~2.0 XLSR-53 front-end~\cite{xlsr53}. Let $z \in
\mathbb{R}^{160}$ be the pre-classification representation. Cross-entropy
alone is the baseline \texttt{CE}; every objective except label smoothing
adds one term,
\begin{equation}
    \mathcal{L} \;=\; \mathcal{L}_{\mathrm{CE}} \;+\;
  \alpha\,\mathcal{L}_{\mathrm{aux}} .
  \label{eq:total}
\end{equation}

\noindent\textbf{Consistency.} The views $x_1 = a_1(x)$ and $x_2 = a_2(x)$ are
independent augmentations of one utterance, compared through a predictor $h$
with $m$ outputs and a stop-gradient on the second, as in
SimSiam~\cite{simsiam}: $\mathcal{L}_{\mathrm{cons}} = D(h(z_1),
\mathrm{sg}[h(z_2)])$, with $D$ the mean-squared (\texttt{MSE}),
mean-absolute (\texttt{MAE}) or cosine (\texttt{cosine}) distance on
$\ell_2$-normalised vectors; the classifier's cross-entropy is computed on
$x_1$ only. For unit vectors $\|a-b\|_2^{2} = 2 - 2\,a^{\!\top}\!b$, and the
mean-squared form also averages over the $m$ coordinates, so
\begin{equation}
  \mathcal{L}_{\mathrm{MSE}} \;=\; \tfrac{2}{m}\,\mathcal{L}_{\cos} .
  \label{eq:equiv}
\end{equation}
With $m = 160$ on AASIST3, \texttt{MSE} at $\alpha = 0.01$ weights
$\mathcal{L}_{\cos}$ by $1.25\times10^{-4}$ against $0.02$ for
\texttt{cosine}: one objective at strengths $160\times$ apart, so
\texttt{MSE} is a near-replicate of \texttt{CE} and serves as a seed-variance
control. The mean-absolute form is not equivalent.

\noindent\textbf{Contrastive.} \texttt{NT-Xent}~\cite{simclr} ($\tau = 0.07$)
on the same views after $h$ and $\ell_2$-normalisation: the other view is the
only positive and every other utterance in the batch a negative, so class
labels are not used.

\noindent\textbf{Adversarial.} Both forms are domain-adversarial
training~\cite{dann} with augmentation as the nuisance: a discriminator $g$
on $z$ predicts whether an utterance is clean or augmented and the encoder is
trained against it, \texttt{GRL} jointly through a reversed gradient and
\texttt{DANN} by alternating updates, the encoder minimising
$\mathcal{L}_{\mathrm{CE}} - \lambda\mathcal{L}_{g}$ while $g$ minimises
$\mathcal{L}_{g}$; $\lambda$ ramps from $0$ to $1$ as in~\cite{dann}.

\noindent\textbf{Label smoothing (\texttt{smooth}).} Cross-entropy with
smoothing $\varepsilon = 0.1$~\cite{mueller2019} on one view, without an
auxiliary term. No setting was tuned: $\alpha = 0.01$ for \texttt{MSE} and
$0.02$ for \texttt{MAE}, \texttt{cosine} and \texttt{NT-Xent}, and a
discriminator of width $64$ for \texttt{GRL} and \texttt{DANN}.

\subsection{Data, training and evaluation}
\label{sec:data}

Within each backbone every run uses identical settings, and no score
normalisation or calibration is applied.

\textbf{1) Data:} Training uses the ASVspoof~5 train split~\cite{asvspoof5},
validated on its dev split. Evaluation covers ASVspoof~5 eval (ASV5), the
only in-domain condition, and four corpora disjoint from training: the
ASVspoof~2021 DeepFake track (ASV21DF)~\cite{asvspoof2021_yamagishi},
In-the-Wild (ITW)~\cite{muller22_generalize},
VoiceWukong (VW)~\cite{voicewukong} and RuASD~\cite{ruasd}.

\textbf{2) Pre-processing:} Audio is resampled to $16$\,kHz, cropped at a
random offset or tiled to $64{,}600$ samples ($\approx 4.04$\,s) and
pre-emphasised.

{\sloppy
\textbf{3) Augmentation:} Every augmented view, diagnostics included, comes
from one \texttt{torch-audiomentations} pipeline: gain ($p{=}0.5$,
$[-10,+6]$\,dB), coloured noise ($0.4$, SNR $[10,40]$\,dB), polarity
inversion ($0.3$), low-pass ($0.3$, $[4,7.5]$\,kHz), high-pass ($0.3$,
$[80,400]$\,Hz), band-pass ($0.2$, $[0.4,3.5]$\,kHz) and band-stop ($0.2$,
$[0.2,4]$\,kHz) filtering, time shift ($0.4$, $\pm10\%$) and pitch shift
($0.25$, $\pm2$ semitones).\par}

\textbf{4) Training AASIST3:} AdamW, $\eta = 10^{-4}$, cosine annealing to
$10^{-6}$, effective batch $48$, $4$ epochs in \texttt{bf16}, the whole model
trained end to end, each of the eight configurations at seeds $1$, $2$ and
$3$. The budget is fixed; Section~\ref{sec:results_collapse} shows the training
classes already separated at it, so evaluation EER reflects unseen attacks
and codecs, not an unfitted model.

\textbf{5) Fine-tuning existing detectors:} The same objectives are applied to
DF-Arena-500M~\cite{dfarena} (DF-Arena below), XLSR-SLS~\cite{xlsr_sls},
Res2TCN~\cite{res2tcn} and WhisperMFCC\allowbreak +\allowbreak
MesoNet~\cite{whispermfcc} (WhisperMFCC below), each with its published
schedule and the backbone at $0.1\times$ the head learning rate ($4$ epochs
at $10^{-4}$ for Res2TCN and WhisperMFCC, $4$ at $5\times10^{-5}$ for
XLSR-SLS, $2$ at $5\times10^{-5}$ for DF-Arena), in three modes: full,
LoRA~\cite{lora} ($r{=}8$, $4$ on WhisperMFCC) and head-only, which trains
the head, the projector $h$ and any discriminator, so the auxiliary term acts
on trainable layers in every mode. In all, $89$ single-seed runs: the $88$ entries
of Table~\ref{tab:ft_modes} and one repeated \texttt{CE} run. Its level, a
mean over five corpora, is not comparable to published in-domain EERs.

\textbf{6) Metrics:} Each run keeps the epoch with the best dev EER. EER is
computed on the pooled scores of each corpus, and the \emph{five-corpus mean}
averages the five, in-domain ASV5 included; AASIST3 EER is averaged over
seeds unless a seed is named. RuASD is near chance for every AASIST3 run and
contributes no order. Scores that are constant or vary only by numerical
noise have no operating point, so their EER is reported as chance, $0.500$
(three DF-Arena runs).

\section{Embedding Diagnostics}
\label{sec:method_diag}

\begin{table*}[!t]
\centering
% In this table the collapsed runs are set in bold (grey elsewhere).
\renewcommand{\coll}[1]{\textbf{#1}}
\small
\setlength{\tabcolsep}{2pt}
\caption{Diagnostics of the $24$ AASIST3 runs on $600$ ASVspoof~5 evaluation utterances, per seed and as mean~$\pm$~sd over seeds (collapsed runs included). Bold marks the four collapsed runs (Section~\ref{sec:results_collapse}): $S$ and silhouette expose them, $A$ ranks them the most invariant and $R$ is unremarkable. Text differences are computed before rounding.}
\label{tab:diag}
\begin{tabular}{l cccc cccc cccc cccc}
\toprule
 & \multicolumn{4}{c}{Displacement $A$} & \multicolumn{4}{c}{Spread $S$} & \multicolumn{4}{c}{Ratio $R$} & \multicolumn{4}{c}{Silhouette} \\
\cmidrule(lr){2-5}\cmidrule(lr){6-9}\cmidrule(lr){10-13}\cmidrule(lr){14-17}
Objective & s1 & s2 & s3 & mean${\pm}$sd & s1 & s2 & s3 & mean${\pm}$sd & s1 & s2 & s3 & mean${\pm}$sd & s1 & s2 & s3 & mean${\pm}$sd \\
\midrule
\texttt{CE} & 2.31 & 1.90 & 4.22 & $2.81{\pm}1.24$ & 5.06 & 3.27 & 6.97 & $5.10{\pm}1.85$ & 0.46 & 0.58 & 0.61 & $0.55{\pm}0.08$ & 0.219 & 0.164 & 0.107 & $0.163{\pm}0.056$ \\
\texttt{MSE} & 1.38 & 2.01 & 4.28 & $2.56{\pm}1.53$ & 3.11 & 3.26 & 8.67 & $5.01{\pm}3.16$ & 0.44 & 0.61 & 0.49 & $0.52{\pm}0.09$ & 0.219 & 0.117 & 0.144 & $0.160{\pm}0.053$ \\
\texttt{MAE} & 3.63 & \coll{0.12} & \coll{0.10} & $1.28{\pm}2.03$ & 9.06 & \coll{0.17} & \coll{0.16} & $3.13{\pm}5.14$ & 0.40 & \coll{0.70} & \coll{0.63} & $0.58{\pm}0.16$ & 0.065 & \coll{0.006} & \coll{0.002} & $0.024{\pm}0.035$ \\
\texttt{cosine} & 1.09 & 1.78 & 2.29 & $1.72{\pm}0.60$ & 2.03 & 3.23 & 3.57 & $2.94{\pm}0.81$ & 0.54 & 0.55 & 0.64 & $0.58{\pm}0.06$ & 0.119 & 0.095 & 0.138 & $0.117{\pm}0.022$ \\
\texttt{NT-Xent} & 1.38 & 2.70 & 3.47 & $2.52{\pm}1.06$ & 1.85 & 5.93 & 8.73 & $5.50{\pm}3.46$ & 0.75 & 0.46 & 0.40 & $0.53{\pm}0.19$ & 0.130 & 0.214 & 0.061 & $0.135{\pm}0.077$ \\
\texttt{smooth} & 3.62 & 2.59 & \coll{0.01} & $2.07{\pm}1.86$ & 5.32 & 4.77 & \coll{0.01} & $3.37{\pm}2.92$ & 0.68 & 0.54 & \coll{0.57} & $0.60{\pm}0.07$ & 0.117 & 0.131 & \coll{0.011} & $0.086{\pm}0.066$ \\
\texttt{GRL} & 3.86 & 2.70 & 8.44 & $5.00{\pm}3.03$ & 5.21 & 4.75 & 13.82 & $7.93{\pm}5.10$ & 0.74 & 0.57 & 0.61 & $0.64{\pm}0.09$ & 0.102 & 0.155 & 0.057 & $0.104{\pm}0.049$ \\
\texttt{DANN} & 3.20 & 2.63 & \coll{0.02} & $1.95{\pm}1.69$ & 5.81 & 5.36 & \coll{0.04} & $3.74{\pm}3.21$ & 0.55 & 0.49 & \coll{0.61} & $0.55{\pm}0.06$ & 0.163 & 0.181 & \coll{0.008} & $0.117{\pm}0.095$ \\
\bottomrule
\end{tabular}
\end{table*}

For each AASIST3 checkpoint we embed a fixed set of $300$ bona fide and $300$
spoof ASVspoof~5 evaluation utterances, each clean and with one augmented
view from the training pipeline (training pairs two augmented views). \emph{Displacement} $A = \mathbb{E}\lVert z -
\tilde{z}\rVert_2$ is the mean distance between an utterance's clean and
augmented representation, and \emph{spread} $S = \mathbb{E}_{i \neq j}\lVert
z_i - z_j\rVert_2$ the mean distance between different utterances. The
Euclidean \emph{silhouette} and the \emph{Fisher ratio}
(distance between the class means over their summed mean per-coordinate
deviations, not the within/between variance ratio of
Section~\ref{sec:results_collapse}) measure class separation, and a
\emph{linear probe} for the class label is a $5$-fold cross-validated
logistic regression, standardised on the training folds.
The \emph{PC1 fraction}, the share of variance on the first principal
component, measures how close the space is to a single direction.

$A$ alone does not measure invariance: replacing $z$ by $cz$ and the
classifier weights $W$ by $W/c$ leaves the scores almost unchanged but scales
$A$ by $c$, so a small $A$ may mean an invariant encoder or merely a small
one. We therefore report $R = A/S$, which is invariant to $c$: alignment over a
uniformity-like spread~\cite{wang2020alignment}, taken in the un-normalised
space the classifier reads rather than on the hypersphere. Normalising $z$
removes $c$ but discards how small the space has become; kept in $S$, a
collapsed run lies $10$--$1000\times$ below the rest, the role of
VICReg's variance term~\cite{bardes2022vicreg}.

We call a run \emph{collapsed} when $S$ falls more than an order of magnitude
below every trained run; its output is near-constant, not constant. A healthy
encoder multiplied by a small constant also has a vanishing $S$, so $S$ only
flags candidates; the scale-free silhouette and Fisher ratio then decide,
since a rescaled encoder keeps both and a collapsed one loses them. $R$ does not detect collapse: a constant output
gives $0/0$, so we leave $R$ undefined at $S = 0$ and read it as invariance
only for runs that trained.

\section{Results}
\label{sec:results}

\begin{figure*}[!t]
  \centering
  \includegraphics[width=0.7\textwidth]{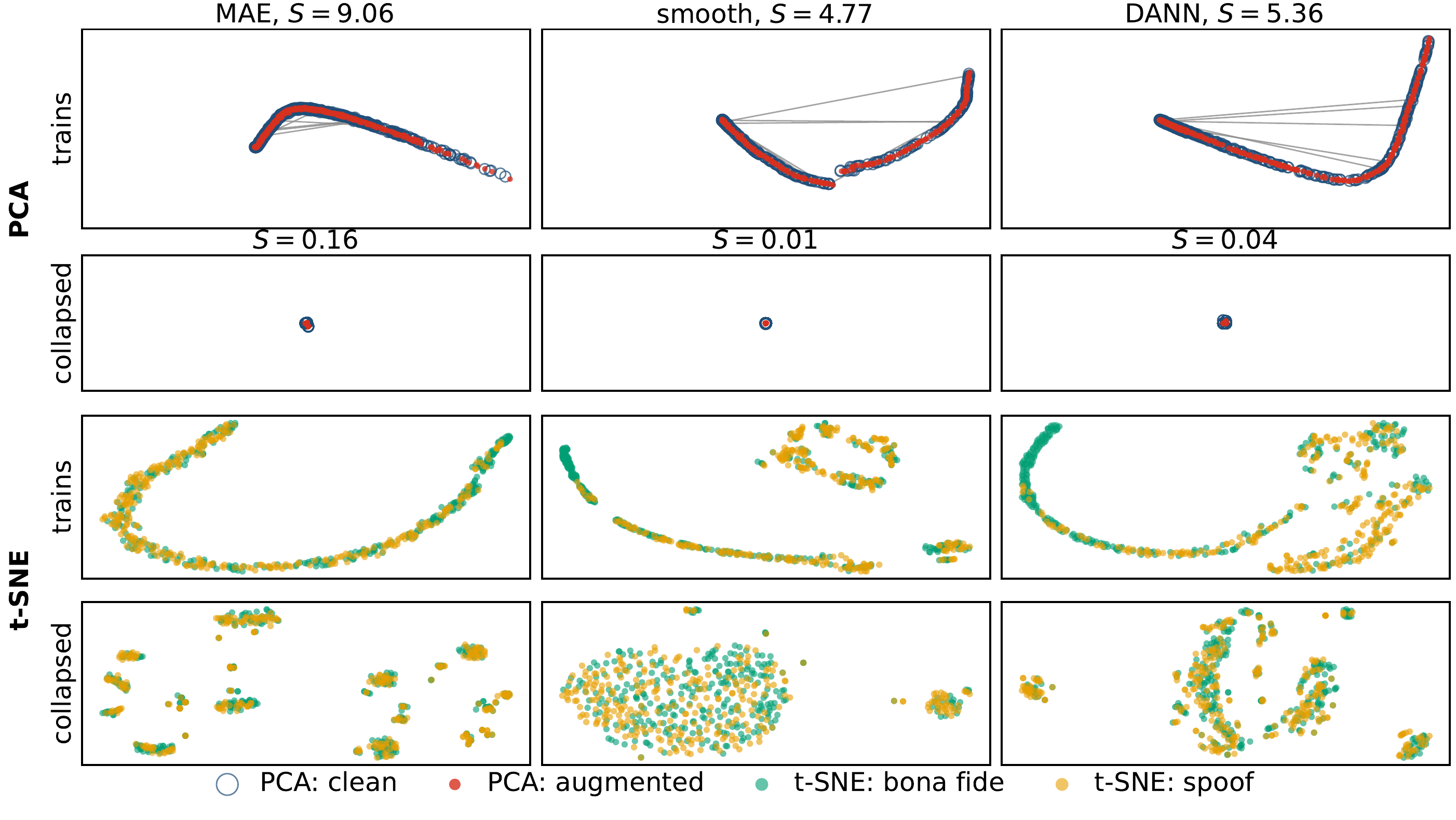}
  \small
  \caption{PCA (top) and t-SNE (bottom) of the $600$ ASVspoof~5 evaluation embeddings for \texttt{MAE}, \texttt{smooth} and \texttt{DANN}, each at a seed that trained and one that collapsed. PCA keeps one axis scale per column, so a trained run is a thin curve and a collapsed run a single point; t-SNE rescales every panel and turns both into similar-looking clusters.}
  \label{fig:space}
\end{figure*}

\begin{table*}[!t]
\centering
\small
\caption{AASIST3 EER per corpus, mean~$\pm$~sd over seeds $1$--$3$ (abbreviations in Section~\ref{sec:data}); the last column averages the five corpora per seed, then over seeds. Bold: lowest in the column (ties share it).}
\label{tab:seed_summary}
\begin{tabular}{l ccccc@{\hspace{10pt}}c}
\toprule
Objective & ASV5 & ASV21DF & ITW & VW & RuASD & 5-corpus mean \\
\midrule
\texttt{CE} & $\mathbf{0.288}{\pm}0.090$ & $0.364{\pm}0.038$ & $\mathbf{0.314}{\pm}0.022$ & $\mathbf{0.293}{\pm}0.026$ & $0.489{\pm}0.033$ & $\mathbf{0.349}{\pm}0.002$ \\
\texttt{MSE} & $0.391{\pm}0.099$ & $0.410{\pm}0.110$ & $0.412{\pm}0.180$ & $0.365{\pm}0.178$ & $0.510{\pm}0.143$ & $0.418{\pm}0.140$ \\
\texttt{MAE} & $0.490{\pm}0.075$ & $0.408{\pm}0.149$ & $0.454{\pm}0.060$ & $0.404{\pm}0.115$ & $\mathbf{0.465}{\pm}0.072$ & $0.444{\pm}0.093$ \\
\texttt{cosine} & $0.383{\pm}0.031$ & $\mathbf{0.354}{\pm}0.008$ & $0.368{\pm}0.032$ & $0.309{\pm}0.077$ & $0.481{\pm}0.037$ & $0.379{\pm}0.023$ \\
\texttt{NT-Xent} & $0.371{\pm}0.043$ & $0.400{\pm}0.068$ & $0.331{\pm}0.068$ & $0.303{\pm}0.089$ & $0.493{\pm}0.083$ & $0.380{\pm}0.065$ \\
\texttt{smooth} & $0.417{\pm}0.097$ & $0.389{\pm}0.077$ & $0.394{\pm}0.116$ & $0.310{\pm}0.170$ & $0.492{\pm}0.093$ & $0.400{\pm}0.096$ \\
\texttt{GRL} & $0.411{\pm}0.033$ & $0.434{\pm}0.103$ & $0.354{\pm}0.015$ & $0.336{\pm}0.118$ & $0.493{\pm}0.060$ & $0.406{\pm}0.062$ \\
\texttt{DANN} & $0.390{\pm}0.127$ & $0.448{\pm}0.091$ & $0.412{\pm}0.109$ & $0.391{\pm}0.139$ & $0.485{\pm}0.036$ & $0.425{\pm}0.091$ \\
\bottomrule
\end{tabular}
\end{table*}

\begin{table*}[!t]
\centering
\small
\caption{Five-corpus mean EER of the fine-tuned detectors, one run each. Bold: lowest in the column (ties share it), an observed order, not a significant difference. Grey: collapsed encoders (Section~\ref{sec:results_collapse}), reported at chance.}
\label{tab:ft_modes}
\begin{tabular}{l ccc ccc ccc ccc}
\toprule
 & \multicolumn{3}{c}{DF-Arena} & \multicolumn{3}{c}{XLSR-SLS} & \multicolumn{3}{c}{Res2TCN} & \multicolumn{3}{c}{WhisperMFCC} \\
\cmidrule(lr){2-4} \cmidrule(lr){5-7} \cmidrule(lr){8-10} \cmidrule(lr){11-13}
Objective & full & LoRA & head & full & LoRA & head & full & LoRA$^{\dagger}$ & head & full & LoRA & head \\
\midrule
\texttt{CE} & 0.454 & 0.488 & 0.570 & \textbf{0.362} & 0.492 & 0.565 & 0.349 & -- & 0.341 & \textbf{0.444} & 0.505 & 0.520 \\
\texttt{MSE} & \textbf{0.375} & 0.564 & 0.567 & 0.500 & 0.498 & 0.568 & 0.374 & -- & 0.333 & 0.541 & 0.513 & 0.527 \\
\texttt{MAE} & 0.417 & 0.567 & 0.565 & 0.372 & 0.487 & 0.562 & 0.363 & -- & 0.331 & 0.508 & 0.506 & 0.493 \\
\texttt{cosine} & 0.383 & 0.566 & 0.566 & 0.387 & 0.517 & 0.566 & 0.376 & -- & 0.343 & 0.528 & 0.511 & \textbf{0.475} \\
\texttt{NT-Xent} & 0.376 & \textbf{0.487} & \textbf{0.562} & 0.538 & 0.470 & 0.560 & 0.364 & -- & \textbf{0.329} & 0.499 & 0.504 & 0.522 \\
\texttt{smooth} & \textbf{0.375} & 0.489 & 0.568 & 0.390 & 0.557 & \textbf{0.432} & \textbf{0.334} & -- & 0.343 & 0.468 & \textbf{0.490} & 0.500 \\
\texttt{GRL} & \coll{0.500} & 0.550 & 0.568 & 0.597 & 0.476 & 0.556 & 0.355 & -- & 0.362 & 0.501 & 0.519 & 0.520 \\
\texttt{DANN} & \coll{0.500} & \coll{0.500} & 0.565 & 0.592 & \textbf{0.456} & 0.554 & 0.335 & -- & 0.350 & 0.510 & 0.526 & 0.526 \\
\bottomrule
\end{tabular}
\par\vspace{2pt}
\parbox{\textwidth}{\small $^{\dagger}$Res2TCN is convolutional and has no attention projection to adapt, so LoRA reduces to head-only.}
\end{table*}

\subsection{Apparent invariance is shrinkage}
\label{sec:results_proxy}

Judged by raw displacement, the quantity usually reported, cosine consistency
appears to work: \texttt{cosine} averages $A = 1.72$ over seeds against
$2.81$ for \texttt{CE}, $39\%$ less. The spread explains the gain. The same
runs average $S = 2.94$ against $5.10$, a space $42\%$ smaller, and the
reduction in $A$ follows the reduction in $S$ seed by seed: on seed~2, where
$S$ is unchanged ($3.23$ against $3.27$), $A$ is unchanged too ($1.78$ against
$1.90$), and on seeds~1 and~3 both fall together. The seed-matched difference
in $R$ between \texttt{cosine} and \texttt{CE} is $+0.08$, $-0.03$ and
$+0.04$, no consistent reduction. Nothing fixes the
scale of $z$ (Section~\ref{sec:method_diag}), so $A$ reports shrinkage as
invariance, and tuning $\alpha$ would change how far the space shrinks, not
this.

Across Table~\ref{tab:diag}, no objective is shown to be more invariant than
cross-entropy: matched by seed, the difference from \texttt{CE} in $R$ changes
sign, or lies within $0.01$ of zero, for every objective with at least two
trained seeds, for \texttt{MSE} $-0.01$, $+0.03$ and $-0.11$ and for
\texttt{NT-Xent} $+0.29$, $-0.13$ and $-0.21$. Effects smaller than the
seed spread of $R$ ($0.06$--$0.19$) cannot be excluded with three seeds. Silhouette agrees: \texttt{NT-Xent} separates the classes less than
\texttt{CE} on two of three seeds.

\subsection{One decision axis in every run, and four collapses}
\label{sec:results_collapse}

Every trained space, \texttt{CE} included, is close to a single direction. In
all $20$ trained runs the first principal component holds $0.63$ to $0.97$ of
the evaluation variance and lies along the difference of the class means
($|\cos| \ge 0.94$). For two classes this is the geometry cross-entropy is
expected to produce~\cite{papyan2020nc}; the finding is that no auxiliary term
alters it: each acts through $h$ or a discriminator, and none constrains the
covariance of $z$ as VICReg does~\cite{bardes2022vicreg}. On $600$ training
utterances the same runs approach neural collapse, with within-class variance
$0.01$ to $0.57$ times the between-class variance and silhouette $0.57$ to
$0.94$, so they are not under-fitted to their training classes. As in image
classification~\cite{hui2022nc}, this structure does not transfer: on
evaluation utterances within-class variance exceeds between-class variance
$3.8$ to $16.5$ times, and the low evaluation silhouette tracks how far the
classes overlap along that axis (Spearman $\rho = 0.94$ with the separation
along it), not how one-dimensional the space is ($\rho = 0.08$ with the PC1
fraction). Four runs go further: their spread vanishes (PC1 fraction
$0.48$--$0.79$, overlapping the trained range).

In these four runs $S$ lies between $0.01$ and $0.17$, in the other $20$
between $1.85$ and $13.82$, with no run in the gap. They come from three
objectives (\texttt{MAE} at two seeds, \texttt{smooth} and \texttt{DANN} at
one), and another seed of each trained normally. Since \texttt{smooth} adds
no term on $z$, no auxiliary term is needed for collapse: it is a failure of
the training recipe that neither the supervised term nor, where
present, SimSiam's stop-gradient and predictor~\cite{simsiam} avert, and that
EER does not reveal. Three of the four collapses occur at seed~3, also the
weakest \texttt{CE} seed (silhouette $0.107$), so the trigger is as plausibly
the seed's trajectory as the objective. They are not rescaled healthy
encoders: silhouette, Fisher ratio and the probe are invariant to a constant
rescaling, yet the flagged runs fall below every trained run in silhouette
($0.002$--$0.011$ against $\ge 0.057$) and Fisher ratio ($0.8$--$2.6$ against
$\ge 3.9$). Collapse is severe, not total:
the probe still finds weak class information (AUC $0.65$--$0.67$). $A$, $R$
and EER miss it: $A$ ranks these runs the most invariant in the study, their
$R$ stays within the $0.40$--$0.75$ of trained runs, and their five-corpus
mean EER of $0.48$--$0.53$ looks like ordinary under-performance. Indeed the
trained \texttt{MSE} run at seed~2 scores $0.575$, worse than any collapsed
run, and \texttt{MAE}, collapsed at two of three seeds, has the lowest RuASD
EER in Table~\ref{tab:seed_summary}, where every objective lies between
$0.47$ and $0.51$. Figure~\ref{fig:space} shows a collapsed run as a single
point at a shared PCA scale, which t-SNE, the usual view, hides. What raw
displacement rewards is thus a small space: the four collapsed runs at the
extreme, and short of it the \texttt{cosine} runs, the smallest trained
spaces.

On DF-Arena three adversarial runs collapse as well (\texttt{GRL} full,
\texttt{DANN} full and LoRA): their encoder output varies by only
$3$--$6\times10^{-4}$ between utterances or not at all, \texttt{DANN} emitting
one score for every input, so Table~\ref{tab:ft_modes} reports them at
chance rather than the tie order of noise.

\subsection{Rankings do not survive a change of seed or architecture}
\label{sec:results_transfer}

A single seed can reverse a ranking. At seed~1 \texttt{MSE} reads $0.307$
five-corpus mean EER against $0.349$ for \texttt{CE}, a $12\%$ gain a
single-run protocol would publish; seed~2 gives $0.575$, the worst of the
$24$ AASIST3 runs. Since \texttt{MSE} at this weight is a near-replicate of
\texttt{CE} (Section~\ref{sec:method_obj}), the two numbers bound the seed
variance of the recipe, and the $0.002$ agreement of the three \texttt{CE}
seeds is a draw. In
Table~\ref{tab:seed_summary} the median over objectives of the seed standard
deviation, $0.064$ to $0.117$ per corpus, exceeds the standard deviation of
the eight objective means, $0.013$ to $0.056$: seeds move EER more than
objectives do. Matched by seed, none has a lower five-corpus mean than
\texttt{CE} on all three seeds, and \texttt{cosine}, \texttt{GRL} and
\texttt{DANN} are higher on every seed.

Table~\ref{tab:ft_modes} is single-seed and cannot rank objectives; it
answers only whether any objective wins everywhere, and none does: under full
fine-tuning the column minimum is \texttt{CE} on two architectures and
\texttt{smooth} on the other two (tied with \texttt{MSE} on DF-Arena), and
\texttt{DANN} is second on Res2TCN and, collapsed, last on DF-Arena. A
repeated \texttt{CE} head-only run on Res2TCN differs by $0.011$, a lower
bound on run-to-run variation since head-only trains the fewest parameters;
within that floor the first two objectives are tied in $6$ of the $11$
columns, and within twice it in $9$. Head-only adaptation narrows the range
relative to full fine-tuning on every architecture, from $0.125$ to $0.008$
on DF-Arena, but does not remove it: head-only \texttt{smooth} is $0.133$
below \texttt{CE} on XLSR-SLS and \texttt{cosine} $0.045$ below it on
WhisperMFCC, candidate effects that need replication.

\section{Discussion and Conclusion}
\label{sec:discussion}

\textbf{Limitations.} All runs train on ASVspoof~5 for at most four epochs
with untuned weights of $0.01$--$0.02$; longer schedules, which bring
AASIST-family detectors with an SSL front-end far lower in-domain EER on
ASVspoof~5~\cite{viakhirev2025scalable}, and larger weights are untested.
Only AASIST3 is repeated, and three seeds show that rankings reverse without
estimating variance or separating seed from objective as the collapse
trigger. Diagnostics cover AASIST3 in domain and the adversarial DF-Arena
runs only, against the training augmentations only; label-aware objectives
(OC-Softmax, SAMO~\cite{zhang2021ocsoftmax,ding2023samo}) are not covered.

A detection metric and a single run cannot tell a shaped space from a shrunken
or collapsed one; under the protocol above no tested objective beats
cross-entropy across seed, architecture and corpus.

\section{Acknowledgments}
The authors used Claude (Anthropic) only to polish the language of the
manuscript; all ideas, experiments, analyses and claims are the authors' own.

\section{Compliance with Ethical Standards}
This study uses only existing, publicly released speech corpora
(Section~\ref{sec:data}); no new human-subject data were collected, and no
ethical approval was required. The authors declare no conflicts of interest.

\bibliographystyle{IEEEbib}
\bibliography{strings,refs}

\begin{thebibliography}{10}

\bibitem{asvspoof2021_yamagishi}
Junichi Yamagishi et~al.,
\newblock ``{ASVspoof} 2021: Accelerating progress in spoofed and deepfake
  speech detection,''
\newblock in {\em Proc. ASVspoof Workshop}, 2021.

\bibitem{muller22_generalize}
Nicolas~M. M{\"u}ller, Pavel Czempin, Franziska Diekmann, Adam Froghyar, and
  Konstantin B{\"o}ttinger,
\newblock ``Does audio deepfake detection generalize?,''
\newblock in {\em Proc. Interspeech}, 2022.

\bibitem{muller24_harder}
Nicolas~M. M{\"u}ller, Nicholas Evans, Hemlata Tak, Philip Sperl, and
  Konstantin B{\"o}ttinger,
\newblock ``Harder or different? understanding generalization of audio deepfake
  detection,''
\newblock in {\em Proc. Interspeech}, 2024.

\bibitem{tak2022}
Hemlata Tak, Massimiliano Todisco, Xin Wang, Jee-weon Jung, Junichi Yamagishi,
  and Nicholas Evans,
\newblock ``Automatic speaker verification spoofing and deepfake detection
  using wav2vec 2.0 and data augmentation,''
\newblock in {\em Proc. Odyssey}, 2022.

\bibitem{viakhirev2025scalable}
Ivan Viakhirev, Daniil Sirota, Aleksandr Smirnov, and Kirill Borodin,
\newblock ``Towards scalable {AASIST}: Refining graph attention for speech
  deepfake detection,''
\newblock {\em arXiv preprint arXiv:2507.11777}, 2025.

\bibitem{clad}
Haolin Wu et~al.,
\newblock ``{CLAD}: Robust audio deepfake detection against manipulation
  attacks with contrastive learning,''
\newblock {\em arXiv preprint arXiv:2404.15854}, 2024.

\bibitem{attn_contrastive}
Chirag Goel, Surya Koppisetti, Ben Colman, Ali Shahriyari, and Gaurav Bharaj,
\newblock ``Towards attention-based contrastive learning for audio spoof
  detection,''
\newblock in {\em Proc. Interspeech}, 2023.

\bibitem{grl_spoof}
Anh-Tuan Dao, Driss Matrouf, Mickael Rouvier, and Nicholas Evans,
\newblock ``Speaker-invariant representation learning for spoofing detection
  via gradient reversal and a variational information bottleneck,''
\newblock {\em arXiv preprint arXiv:2606.08678}, 2026.

\bibitem{simsiam}
Xinlei Chen and Kaiming He,
\newblock ``Exploring simple siamese representation learning,''
\newblock in {\em Proc. CVPR}, 2021.

\bibitem{simclr}
Ting Chen, Simon Kornblith, Mohammad Norouzi, and Geoffrey Hinton,
\newblock ``A simple framework for contrastive learning of visual
  representations,''
\newblock in {\em Proc. ICML}, 2020.

\bibitem{dann}
Yaroslav Ganin and Victor Lempitsky,
\newblock ``Unsupervised domain adaptation by backpropagation,''
\newblock in {\em Proc. ICML}, 2015.

\bibitem{mueller2019}
Rafael M{\"u}ller, Simon Kornblith, and Geoffrey Hinton,
\newblock ``When does label smoothing help?,''
\newblock in {\em Proc. NeurIPS}, 2019.

\bibitem{zhang2021ocsoftmax}
You Zhang, Fei Jiang, and Zhiyao Duan,
\newblock ``One-class learning towards synthetic voice spoofing detection,''
\newblock {\em IEEE Signal Processing Letters}, vol. 28, pp. 937--941, 2021.

\bibitem{ding2023samo}
Siwen Ding, You Zhang, and Zhiyao Duan,
\newblock ``{SAMO}: Speaker attractor multi-center one-class learning for voice
  anti-spoofing,''
\newblock in {\em Proc. ICASSP}, 2023.

\bibitem{rawboost}
Hemlata Tak, Madhu Kamble, Jose Patino, Massimiliano Todisco, and Nicholas
  Evans,
\newblock ``{RawBoost}: A raw data boosting and augmentation method applied to
  automatic speaker verification anti-spoofing,''
\newblock in {\em Proc. ICASSP}, 2022.

\bibitem{viakhirev2026interpret}
Ivan Viakhirev, Kirill Borodin, Mikhail Gorodnichev, and Grach Mkrtchian,
\newblock ``Interpreting multi-branch anti-spoofing architectures: Correlating
  internal strategy with empirical performance,''
\newblock {\em Mathematics}, vol. 14, no. 2, pp. 381, 2026.

\bibitem{bouthillier2021variance}
Xavier Bouthillier et~al.,
\newblock ``Accounting for variance in machine learning benchmarks,''
\newblock in {\em Proc. MLSys}, 2021.

\bibitem{papyan2020nc}
Vardan Papyan, X.~Y. Han, and David~L. Donoho,
\newblock ``Prevalence of neural collapse during the terminal phase of deep
  learning training,''
\newblock {\em Proceedings of the National Academy of Sciences}, vol. 117, no.
  40, pp. 24652--24663, 2020.

\bibitem{aasist3}
Kirill Borodin et~al.,
\newblock ``{AASIST3}: {KAN}-enhanced {AASIST} speech deepfake detection using
  {SSL} features and additional regularization for the {ASVspoof} 2024
  challenge,''
\newblock in {\em Proc. ASVspoof Workshop}, 2024.

\bibitem{xlsr53}
Alexis Conneau, Alexei Baevski, Ronan Collobert, Abdelrahman Mohamed, and
  Michael Auli,
\newblock ``Unsupervised cross-lingual representation learning for speech
  recognition,''
\newblock in {\em Proc. Interspeech}, 2021.

\bibitem{asvspoof5}
Xin Wang et~al.,
\newblock ``{ASVspoof} 5: Crowdsourced speech data, deepfakes, and adversarial
  attacks at scale,''
\newblock in {\em Proc. ASVspoof Workshop}, 2024.

\bibitem{voicewukong}
Ziwei Yan, Yanjie Zhao, and Haoyu Wang,
\newblock ``{VoiceWukong}: Benchmarking deepfake voice detection,''
\newblock {\em arXiv preprint arXiv:2409.06348}, 2024.

\bibitem{ruasd}
Ksenia Lysikova, Kirill Borodin, and Grach Mkrtchian,
\newblock ``Evaluating generalization and robustness in {R}ussian
  anti-spoofing: The {RuASD} initiative,''
\newblock {\em IEEE Access}, vol. 14, pp. 102394--102410, 2026.

\bibitem{dfarena}
Ajinkya Kulkarni, Sandipana Dowerah, Atharva Kulkarni, Tanel Alum{\"a}e, and
  Mathew Magimai~Doss,
\newblock ``Do compact {SSL} backbones matter for audio deepfake detection? {A}
  controlled study with {RAPTOR},'' 2026.

\bibitem{xlsr_sls}
Qishan Zhang, Shuangbing Wen, and Tao Hu,
\newblock ``Audio deepfake detection with self-supervised {XLS-R} and {SLS}
  classifier,''
\newblock in {\em Proc. ACM Multimedia}, 2024, pp. 6765--6773.

\bibitem{res2tcn}
Kirill Borodin, Vasiliy Kudryavtsev, Grach Mkrtchian, and Mikhail Gorodnichev,
\newblock ``Capsule-based and {TCN}-based approaches for spoofing detection in
  voice biometry,''
\newblock {\em Engineering, Technology \& Applied Science Research}, vol. 14,
  no. 6, pp. 18409--18414, 2024.

\bibitem{whispermfcc}
Piotr Kawa, Marcin Plata, Micha{\l} Czuba, Piotr Szyma{\'n}ski, and Piotr Syga,
\newblock ``Improved deepfake detection using whisper features,''
\newblock in {\em Proc. Interspeech}, 2023, pp. 4009--4013.

\bibitem{lora}
Edward~J. Hu et~al.,
\newblock ``{LoRA}: Low-rank adaptation of large language models,''
\newblock in {\em Proc. ICLR}, 2022.

\bibitem{wang2020alignment}
Tongzhou Wang and Phillip Isola,
\newblock ``Understanding contrastive representation learning through alignment
  and uniformity on the hypersphere,''
\newblock in {\em Proc. ICML}, 2020.

\bibitem{bardes2022vicreg}
Adrien Bardes, Jean Ponce, and Yann LeCun,
\newblock ``{VICReg}: Variance-invariance-covariance regularization for
  self-supervised learning,''
\newblock in {\em Proc. ICLR}, 2022.

\bibitem{hui2022nc}
Like Hui, Mikhail Belkin, and Preetum Nakkiran,
\newblock ``Limitations of neural collapse for understanding generalization in
  deep learning,''
\newblock {\em arXiv preprint arXiv:2202.08384}, 2022.

\end{thebibliography}

\end{document}